\documentclass[runningheads]{llncs}

\usepackage{amsmath}
\usepackage{algorithmic}
    
\usepackage{graphicx}
\usepackage{longtable}
\usepackage{amsfonts}
\usepackage{amssymb}
\usepackage{bm}
\usepackage[utf8]{inputenc}
\usepackage[english]{babel}
\usepackage{xstring}
\usepackage{setspace}
\usepackage{array}
\usepackage{subfig}
\usepackage{enumitem}
\usepackage{multirow}
\usepackage{algorithmic}
\usepackage{dutchcal} 
\usepackage{ctable} 
\usepackage{comment}
\usepackage[ruled,vlined]{algorithm2e}
\usepackage{balance}
\usepackage{rotating}
\usepackage{url}

\usepackage{calligra}

\newcolumntype{M}[1]{>{\centering\arraybackslash}m{#1}}
\newcolumntype{N}{@{}m{0pt}@{}}

\font\mysubtitlefont=cmr12 at 12pt

\begin{document}

\raggedbottom
\title{What was Said, What was not Said\\}
\subtitle{\mysubtitlefont Challenges and Good Practices in\\Safety Requirements Specification (SRS)}
\author{Hamid Jahanian}
\institute{FS Expert (TÜV Rheinland) \#266/16-SIS\\UGL, Sydney, Australia\thanks{This article represents the author's own view, and not the position of his employer.}\\ 
\email{hamid.jahanian@ugllimited.com}\\
\url{linkedin.com/in/hamid-jahanian}}

\maketitle

\section{Introduction}\label{Sec_Introduction}

In the process industry, the configuration of Safety Instrumented Systems (SIS) must comply with a defined set of safety requirements, typically documented in the Safety Requirements Specification (SRS). The functional safety standard IEC 61511 outlines the necessary content and quality criteria for the SRS. However, developing an effective SRS can be challenging.

This article examines some of these challenges and proposes good practices to address them. It discusses SRS ownership, staged development of SRS, and the classification and traceability of requirements. Additionally, it explores the issue of untold ``negative'' requirements and suggests exploratory ``inspection'' of SIS Application Programs (APs) as a potential remedy.

Some of the ideas presented here are inspired by general systems engineering concepts, though the good practices given here are not strictly tied to systems engineering methodologies.

While the article primarily focuses on the process industry, its core principles can be applied to other sectors as well.

This article is intended for engineers with experience in functional safety and, therefor, does not provide introductory background information. Readers seeking additional context can refer to the sources listed at the end.

\section{What was said (specified requirements)}\label{Sec_ReqDef}

According to systems engineering standards \cite{Ref_345,Ref_346}, a requirement is a ``statement which translates or expresses a need and its associated constraints and conditions.”

Functional safety standards \cite{Ref_190,Ref_186} use different terminology to convey the same concept. IEC 61511 defines the SRS as a ``specification containing the functional\footnote{The term ‘functional’ here does not limit the SRS scope to only functional requirements. IEC 61511 also mandates the inclusion of other requirement types (e.g., environmental); see Clause 10.3.2 in \cite{Ref_190}.} requirements for the SIFs [Safety Instrumented Functions] and their associated safety integrity levels [SILs].”

The purpose of specifying requirements is to establish the constraints and conditions necessary for ensuring the safe operation of the plant during its operational phase. However, safety requirements are primarily addressed during the design and development stages. As stated in Clause 10.3.2 of IEC 61511 \cite{Ref_190}: ``These requirements shall be sufficient to design the SIS and shall include a description of the intent and approach applied during the development of the SIS safety requirements, as applicable.” While some requirements overlap with operation and maintenance (e.g., bypassing and proof testing), the SRS predominantly focuses on design and development. Requirements solely related to operation and maintenance (e.g., operator training) are covered in other sections of the standard, such as Section 16.

\section{Two challenges with one remedy}

Both systems engineering standards and functional safety standards adopt a life cycle approach. These standards define the necessary processes for different life cycle stages, including the steps required for specifying requirements.

In basic systems engineering, requirement definition follows a two-stage process: the definition of stakeholder needs and the definition of system requirements \cite{Ref_345,Ref_346,Ref_348,Ref_347}. The first stage establishes high-level conditions and constraints critical for meeting user needs. The second stage translates these high-level definitions into more technical, measurable requirements that the system must fulfill.

In contrast, IEC 61511 does not specify a staged approach. However, it does define the timing by requiring that the SRS be derived from the allocation of SILs in the preceding life cycle stage and be developed before the design stage (Clause 10.2 and Figure 7 in \cite{Ref_190}).

In practice, not every detail required for the final design and implementation needs to be established before design begins, nor is it always possible to define such requirements too early in a project. Design and implementation are often the longest phases of an integration project and typically undergo multiple iterations before final decisions are made. Expecting a fully complete and exhaustive SRS too early can be impractical.

A more pragmatic alternative is a ``staged'' approach. The initial revision of the SRS can cover the minimum set of requirements necessary for conceptual SIS design (i.e., basic stakeholder needs), with additional requirements incorporated as the design progresses. Each SRS element must still be established before it is needed for the design—but only at that stage, not from the outset.

This progressive development of the SRS does not mean that the SRS follows the design. Rather, the design continues to follow the SRS, but in a stepwise manner rather than all at once. By adopting this approach, not only are the standard’s requirements met in terms of developing an SRS and an SRS-based design, but the resulting SRS is also more likely to contain relevant and effective information, as it evolves progressively, incorporating insights gained from preceding steps. The remaining question will then be how to structure the SRS elements to align with the stages of project. We address this in the next section.

~

A second challenge in SRS development concerns ownership and responsibility. Unlike safety standards such as EN 50126 \cite{Ref_349}, which is relatively prescriptive in defining roles and responsibilities, IEC 61511 does not assign responsibility for developing the SRS to a specific role or organization. Arguably, such a designation would be impractical for a performance-based standard.

Nevertheless, in reality, a well-structured SRS requires input from both the plant owner and the supplier. On one hand, the SRS addresses operational constraints and conditions, meaning it should conceptually be owned and developed by the plant owner, who is best positioned to define these constraints. On the other hand, specifying detailed, technology-specific requirements also requires input from the supplier.

As noted earlier, systems engineering addresses this challenge by dividing requirement specification into two parts: stakeholder needs and system requirements. Breaking down requirement specification into smaller steps—whether two or more—can help distribute ownership effectively. In the initial stages, ownership primarily rests with the plant owner, who specifies needs and high-level requirements. As the SRS development progresses, responsibility shifts toward the system supplier, who defines detailed technical requirements.

\section{Divide and credit}

In the staged approach, the SRS document is developed incrementally, allowing each section to be completed as soon as the prerequisite information becomes available and before it is needed for design or development. This approach is inspired by the two-stage requirements specification and the Agile methodology in systems engineering \cite{Ref_345,Ref_347}.

\begin{figure}[!h] \begin{center} \includegraphics[scale=0.5]{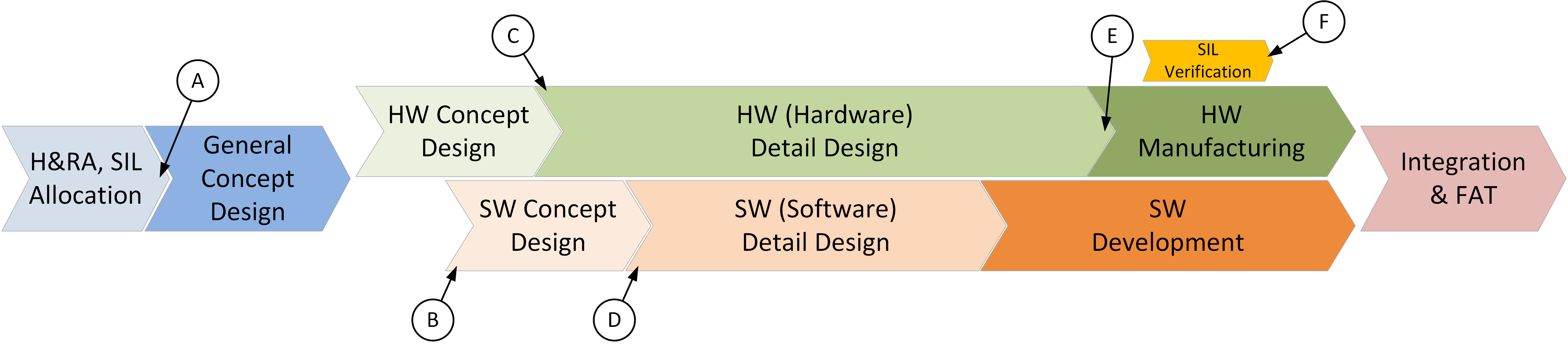} \end{center} \caption{Partial Project Timeline} \label{Fig_Prog} \end{figure}

We illustrate this process with an example. Consider a typical SIS project, for which a partial timeline is shown in Figure \ref{Fig_Prog}. In this example, a combined SRS and AP SRS is developed in six steps, labeled here by letters A to F.

Clauses 10.3.2 and 10.3.5 of IEC 61511 define the content requirements for the SRS and AP SRS, respectively. In this example, we group these requirements into Steps A to F as follows:

\begin{enumerate}[label=Step \Alph*: , leftmargin=1.32cm] \item Following the Hazard and Risk Assessment (H\&RA) and SIL allocation, the first revision of the SRS is established. This includes defining all SIFs, their safety functions, safe states, demand rates, response times, SIL requirements, spurious trip limits, and common cause failure considerations.
\item Before software (SW) design begins, general AP requirements are added, including the list of SIFs and their SILs, real-time performance criteria, operator interfaces, process operation modes, and reference documents.
\item Prior to detailed hardware (HW) design, relevant requirements are incorporated. These include detailed specifications of SIF-related plant I/O devices, process measurements, output actions, manual shutdown requirements, trip mechanisms, system interfaces, hazardous output states, environmental conditions, process operating modes, accident survival requirements, and proof test hardware requirements.
\item Before detailed SW design, software-related requirements are specified. These include hazards from concurrent safe states, SIF input-output logic, reset requirements, failure modes and SIS responses, SIF operation across plant modes, and proof test SW requirements. Additionally, AP requirements are defined, covering program sequencing, handling of bad process variables, proof testing and diagnostics for external devices, application self-monitoring, device monitoring, periodic SIF testing, communication interface requirements, identification and avoidance of dangerous process states, and process variable validation criteria.
\item Upon completion of detailed HW design, requirements for feasible SIS repair times and specific startup/restart procedures are documented.
\item Once SIL verification is complete, proof test intervals and implementation details, bypass requirements, actions to maintain a safe state in the event of SIS faults, and Common Cause Failure (CCF) considerations are finalized.
\end{enumerate}

These six requirement groups will be developed in six stages, as shown in Figure \ref{Fig_Prog}: A, E, and F at the end of the life cycle activity when they can be developed, and B, C, and D beforehand, when they are needed. 

~

\begin{figure}[!h] \begin{center} \includegraphics[scale=0.54]{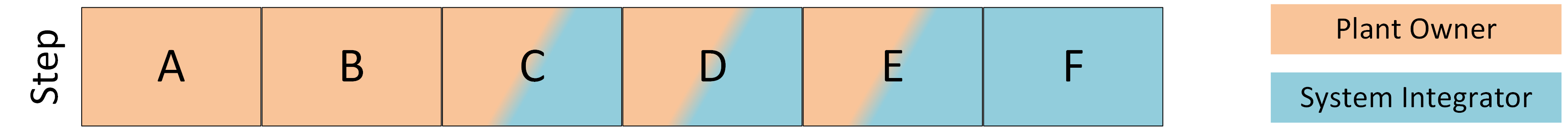} \end{center} \caption{Requirements Ownership} \label{Fig_Own} \end{figure}

Ownership of the SRS can be shared among different parties, depending on the nature of the requirements. Typically, ownership starts with the Plant Owner (PO) and gradually transitions to the System Integrator (SI). In this example, and based on the requirement groupings outlined above, ownership distribution will be as illustrated in Figure \ref{Fig_Own}. 

Regardless of primary ownership, the SI’s established requirements should always be reviewed and confirmed by the PO, and those set by PO should always be checked by SI before implementation. 

Also, in real-world projects, additional stakeholders (e.g., subcontractors of both the PO and SI) may be involved, requiring further refinement of ownership and/or allocation of requirements.

\section{More good practices}

An SIS consists of hardware components and software elements. It may interface with third-party systems and, almost always, with the plant operator. It performs SIFs, which are assigned target reliability measures. At a later stage, the SIS undergoes testing, operation, maintenance, and modification. 

Any requirements set for the SIS should relate to one of the aforementioned aspects, and the SRS document should include all such requirements. To facilitate this, we recommend classifying the requirements based on their primary role. An example of such classifications is as follows:

\begin{enumerate}
\item Functional (FU): Defines the expected function of the SIS and SIFs in different modes of operation.
\item Reliability (RE): Includes target reliability measures and target SILs.
\item Hardware (HW): Specifies constraints and conditions related to SIS components, including EMC and environmental restrictions.
\item Software (SW): Details the software requirements of the SIS.
\item Operational (OP): Outlines what the operation and maintenance crew need to do during the operation phase. Some of these requirements should also be considered during the design stage to ensure that necessary provisions are made to facilitate future operation and maintenance.
\item Compliance (CM): Covers requirements such as conformance to specific safety standards and/or documentation systems.
\item Interface (IN): Includes security and third-party interface requirements.
\end{enumerate}

Classifying requirements helps in clarifying their nature and purpose. In addition, when taking the staged approach, classification also helps in timing the specification of requirements in relation to the project timeline.

~

Furthermore, we recommend assigning traceability tags to specific requirements in the SRS document. This identification helps distinguish between the normative and informative parts of the document while maintaining continuity within the SRS text. More importantly, unique identification tags facilitate requirement tracking during verification and validation processes.

An example of such traceability tags is as follows: \textit{``\textcolor{blue}{[ID: R0503, Cat: RE]} Safety Function SIF01 must meet a minimum SIL rating of 2.''} Here, the first part of the tag provides a unique identification number, while the second element indicates the requirement class as defined earlier.

~

Some projects use requirement management systems, such as IBM DOORS \cite{Ref_352}, to manage all project requirements, including safety, in a database. The use of such tools is essential and beneficial, particularly in large-scale projects. However, in the author's view, having a well-structured SRS document offers significant advantages. 

Firstly, a document provides contextual information that enhances the understanding of the objective and background of requirements. In contrast, databases often isolate requirements from their supporting background.

Secondly, when managed as isolated database entries, requirements typically need to be expressed in SMART (Specific, Measurable, Achievable, Relevant, Time-bound) form \cite{Ref_351} to enable the verification and validation of individual requirements. The challenge is that many safety requirements that are based on performance-based standards such as IEC 61508 and IEC 61511 are inherently subjective and relative, making them difficult to frame in a strict SMART format.

For example, the SRS may require that \textit{``CCF must be reduced to as low as practically possible.''} This is a necessary requirement, but it is not objectively measurable and, unlike some other SMART requirements, cannot be objectively verified. Subjective requirements like this are better assessed through judgment-based evaluations, such as a Functional Safety Assessment (FSA), where the assessor can check the design documents and make an expert judgment.

It is also important to distinguish between the subjectivity of a requirement and the clarity of its expression in the SRS. As mandated by IEC 61511 (Clause 10.2 of \cite{Ref_190}), all requirements should be clear, precise, verifiable, maintainable, and feasible. The fact that a requirement is not measurable does not mean it cannot be clearly expressed. However, its verification may still require subjective methods, such as FSA.

\section{What was not said (untold requirements)}

SRS documents typically specify what the system is supposed to do, not what it is \emph{not} supposed to do. For example, it is common to see a requirement like: \textit{``A close command must be initiated to the valve if the tank pressure exceeds the 98bar threshold."} However, it is far less common to see statements like: \textit{``No alarms should stay active on the operator screen when the plant is in ‘normal’ operation mode''} or \textit{``No reset command should be initiated to energize the trip relay while the pump is in an emergency shutdown state."}

Requirements like the first example above are ``positive'' requirements, specifying actions that must occur under certain conditions. The other two statements are ``negative'' requirements, specifying something that must \emph{not} happen under certain conditions.

The use of negative requirements is more prevalent in some other disciplines. For example, in software development, the negative requirement: \textit{``The program must not enter an infinite loop"} is a common safety requirement. However, in the author's experience, SRS documents in the process industry are typically written in positive language.

Positive requirements typically define the system’s intended functions, whereas negative requirements specify undesired system faults. One reason negative requirements are often overlooked is the assumption that system faults fall under the responsibility of system suppliers, who are expected to know them from experience and address them accordingly. 

Obviously, this assumption is not universally valid, for two reasons: First, not all system faults originate from generic system components. Some faults are project-specific and should be addressed at the project level (e.g., errors in the AP). Second, project engineers have diverse backgrounds and experiences, which may not always align with the specific needs of the project at hand.

Regardless of the cause, excluding negative requirements increases the risk of systematic faults. Such faults may remain dormant in the system, only to surface later during testing and commissioning—or worse, during actual operation.

To be clear, a negative requirement is not simply the negation of a positive requirement, and we are certainly not advocating redundant information in the SRS. Negative requirements are standalone requirements with one distinctive quality: they require the \emph{absence} of an event rather than its presence. What we are advocating is ensuring that all foreseeable failure scenarios—both positive and negative—are examined and, if necessary, included in the SRS as requirements. The goal is to identify and address potential systematic faults early in the design, to ensure they will be fixed and validated prior to operation. 

Methods such as FMEA \cite{Ref_181} can help systematically identify negative failure scenarios. Alternatively, taking a progressive approach to SRS development can help ensure the pragmatic inclusion of negative requirements, as these less obvious requirements may become more apparent as the design progresses.

\section{Dangerous Unrequired}

The SRS is the cornerstone of any SIS project. All activities that impact the integrity of the SIS—including design specification, design, development, verification, and validation—directly or indirectly rely on this critical document. The biggest challenge, therefore, is ensuring the correctness and completeness of the SRS itself. If a necessary requirement is not captured in the SRS, it will not be covered in the design and, consequently, will not be examined during verification and validation.

To address this gap in relation to negative requirements, let us take a closer look at the source of the problem. As mentioned earlier, negative requirements primarily reflect system faults, as opposed to system functionalities. The sources of system faults generally fall into the following categories:
\begin{enumerate}
\item Hardware (HW) faults at the component level, including failures in complex parts such as the Logic Solver.
\item HW faults at the system integration level (i.e., system architectural).
\item Software (SW) faults at the component level (e.g., embedded software).
\item System-level SW faults, such as errors in the Logic Solver’s operating system.
\item Application-level SW faults (i.e., the SIS AP).
\end{enumerate}

Among these categories, Items 2 and 5 are where the system integrator has the greatest influence—and, consequently, the highest likelihood of introducing errors during system integration. Between these, Item 5 (SIS AP) is particularly critical because:  a) Software configuration is inherently more complex than hardware architecture, increasing the risk of unknown failure scenarios; and b) Software is easier to modify than hardware, making it more susceptible to errors introduced through changes.

Given the scope of this article, we focus on one source of systematic faults: the SIS AP, which is, nonetheless, the most critical source of faults in system integration projects.

Standard assurance processes, such as verification and validation, can detect errors in SIS APs. However, as mentioned earlier, these conformity processes primarily assess compliance with the SRS, which may only cover positive requirements. Where including negative requirements in the SRS is impractical or undesirable, a detailed exploratory ``inspection'' of the SIS AP is necessary to uncover all possible failure scenarios—whether documented in the SRS or not.

The challenge of a by-hand exploratory inspection lies in the complexity of industrial SIS APs. It is often impractical for human analysts to manually navigate all pathways within the AP and identify potential failure behaviors. 

Failure Mode Reasoning (FMR) \cite{Ref_350,Ref_225_0} was developed to address this challenge. FMR analyzes the actual SIS AP code and examines how faulty behaviors in SIS inputs or program parameters could lead to undesired state changes at the outputs. This analysis is based on a logical reasoning process, evaluating local failure behaviors and the interdependencies between local functions, to determine global failure behaviors. 

By examining the actual SIS AP code, FMR assesses the system as it exists, independently of any documented requirements. While the primary goal of FMR is not to modify the SRS, its comprehensive failure analysis can enhance understanding of system behavior and help identify potential gaps in the SRS.

Another method for conducting an exploratory inspection of the SIS AP is automated SIS simulation testing. In this approach, a test software interfaces with the SIS through its SW inputs and outputs, running and recording automatically generated test cases.

Manually creating and executing test cases is a labor-intensive process, typically limited to what is required by the SRS. In contrast, automated testing provides a wider test coverage, that can help identify unexpected failure behaviors and potential gaps in the SRS.

In the author’s experience, both FMR and automated test case generation have proven valuable in uncovering failure scenarios not covered by the SRS. However, we do recognize that implementing these additional measures often requires custom tool development, as commercial off-the-shelf solutions for such advanced analyses are generally unavailable.

\section{Conclusion}

We highlighted some common challenges related to the SRS and proposed remedies to address them. We discussed good practices, including a staged approach to SRS development and exploratory investigations of SIS AP. We do not claim that the good practices outlined in this article provide a one-size-fits-all solution. Nor do we suggest that they completely eliminate systematic failures related to the SRS. However, we believe that they can significantly reduce the likelihood of systematic failures by improving clarity and accuracy.

In summary, our recommendations are as follows: classify the requirements, assign traceability tags to specific requirements, use these tags to distinguish between normative and informative parts of the SRS, consider the inclusion of negative requirements, agree with stakeholders on responsibility division for defining requirements, develop the SRS progressively throughout the design stage, and consider using exploratory inspection of APs in addition to the usual conformity-based verification practices.

\bibliographystyle{splncs04}
\bibliography{References1}

\end{document}